\documentclass[conference]{IEEEtran}

\usepackage[T1]{fontenc}
\usepackage[utf8]{inputenc}
\usepackage{upquote}
\def\IEEEbibitemsep{3pt plus .5pt}

\usepackage{graphics}
\usepackage{tikz}

\usepackage{tikz}
\usepackage{graphicx}
\usepackage{bm}
\graphicspath{{./results/plots/}}
\usepackage{subfloat}
\usepackage{color}
\usepackage{comment}
\usepackage{amsmath}
\usepackage{balance}

\usepackage{amsfonts}
\usepackage[normalem]{ulem}
\usepackage{pbox}
\usepackage[hidelinks]{hyperref}
\usepackage{tabularx}
\usepackage{multirow}
\usepackage{booktabs}
\usepackage{multicol}
\usepackage{makecell}
\usepackage[caption=false]{subfig}
\usepackage{mathtools}

\usepackage{authblk}

\usepackage[vlined,ruled]{algorithm2e}

\usepackage{color}
\usepackage{array}
\newcolumntype{L}[1]{>{\raggedright\let\newline\\\arraybackslash\hspace{0pt}}m{#1}}
\newcolumntype{C}[1]{>{\centering\let\newline\\\arraybackslash\hspace{0pt}}m{#1}}
\newcolumntype{R}[1]{>{\raggedleft\let\newline\\\arraybackslash\hspace{0pt}}m{#1}}
\usepackage{listings}
\usepackage{adjustbox}

\usepackage{xcolor}
\begin{document}

\title{Combining Performance and Productivity: Accelerating the Network Sensing Graph Challenge with GPUs and Commodity Data Science Software}

\bstctlcite{IEEEexample:BSTcontrol}

\author{Siddharth Samsi}
\author{Dan Campbell}
\author{Emanuel Scoullos}
\author{Oded Green}

\affil{NVIDIA}

\maketitle
\thispagestyle{plain}
\pagestyle{plain}

\begin{abstract}

The HPEC Graph Challenge is a collection of benchmarks representing complex workloads that test the hardware and software components of HPC systems, which traditional benchmarks, such as LINPACK, do not. The first benchmark,  Subgraph Isomorphism, focused on several compute-bound and memory-bound kernels. The most recent of the challenges, the Anonymized Network Sensing Graph Challenge, represents a shift in direction, as it represents a longer end-to-end workload that requires many more software components, including, but not limited to, data I/O, data structures for representing graph data, and a wide range of functions for data preparation and network analysis. A notable feature of this new graph challenge is the use of GraphBLAS to represent the computational aspects of the problem statement. In this paper, we show an alternative interpretation of the GraphBLAS formulations using the language of data science. With this formulation, we show that the new graph challenge can be implemented using off-the-shelf ETL tools available in open-source, enterprise software such as NVIDIA's RAPIDS ecosystem. Using off-the-shelf software, RAPIDS cuDF and cupy, we enable significant software acceleration without requiring any specific HPC code and show speedups, over the same code running with Pandas on the CPU, of 147x-509x on an NVIDIA A100 GPU, 243x-1269X for an NVIDIA H100 GPU, and 332X-2185X for an NVIDIA H200 GPU.

\end{abstract}

\section{Introduction}
\label{sec:intro}
The release of large, real, and diverse data sets to the public allows researchers to inspect and analyze the data more thoroughly, extract meaningful insights, and test the performance and scalability of algorithms. 
The MNIST \cite{lecun1998mnist} and ImageNet \cite{deng2009imagenet} datasets with large numbers of labeled images have enabled researchers to create new ways to label and classify images. 
The TPC-H benchmark (\texttt{\url{https://www.tpc.org/tpch/}}) and its respective datasets have enabled researchers to create new database solutions that target different computational workloads. 
The MIT/Amazon/IEEE Graph Challenge (\texttt{\url{https://graphchallenge.mit.edu}}) has a collection of graph processing problems that can be tested on a mix of real and synthetic data sets.

The graph research community is split between the use of synthetic data (the RMAT \cite{Rmat2004} format used in the Graph500 benchmark \cite{graph500}) and real-world datasets (such as those made available in the SuiteSparse collection \cite{davis2011university} and SNAP \cite{Snap-Stanford}). 
Synthetic data is used for testing the scalability of a solution. However, in most cases, it cannot be used to qualitatively analyze a real-world problem or domain because it cannot match real-world data in every characteristic. 
Real-world datasets would address this, but they suffer from numerous limitations such as: the matrices are too small, they may be derived from an area of computational science that may not fully reflect the properties of the target domain or the matrices underwent a cleansing process before their release. The last of these is especially problematic when the research focuses on anomaly detection or finding finer-grain properties of the network. 

Owners of raw data face their own challenges in releasing the data. For example, the raw data might contain private information that cannot be disclosed. By anonymizing the data and removing the ability to reverse engineer the true source of the data, privacy concerns can be addressed, and the data can be released. 

The Anonymized Network Sensing graph challenge focuses on the anonymization of Packet Capture (PCAP) data.
While PCAP data is unique to communication networks, a significant part of the graph challenge workload can be extended to other types of networks.  Thus, the goal of the newly created benchmark by Jananthan \emph{et al.} \cite{hayden2024} is to show that these challenges can be met to release to the public a broader (and possibly uncleaned) datasets that will enable the development of new network science tools. Additionally, loading PCAP data brings its own set of challenges, namely the sequential nature of the binary representation. We cover this in additional detail below and note that using more data-friendly formats, such as Parquet or CSV (Comma Separated Values) when storing data to disk, is one way to make this graph challenge friendlier to the data-science community.

\subsection*{Contributions}
$\bullet$ In this paper, we show a simple solution that meets the new benchmark requirements and is entirely based on commodity hardware (NVIDIA GPUs) and commodity open-source software. Our code is based on the RAPIDS ecosystem and uses cuDF~\cite{cudf} for data analytics and open source libraries \texttt{dkpt}\cite{dpkt} and GraphBLAS\cite{davis2019algorithm} for parsing data provided as part of this challenge. 
Our goal when designing our solution was to provide a turnkey solution that would not require implementing or optimizing CUDA kernels, as we believe most data scientists rely on established tools when implementing their pipelines. Our secondary goal was to ensure that our pipeline was balanced, without using very computationally intensive functions, while also being certain that results were correct. Both goals are achieved. 

$\bullet$ We take a data science approach to solving the new graph challenge by making the solution available to a broad audience. We believe that our new solution will effectively complement the reference GraphBLAS-based solution. We note that while the problem statement does not state a solution formulation, using GraphBLAS-based solutions allows for correctness verification with relative ease. By implementing a non-trivial solution and showing the relationship between certain GraphBLAS operations and data science functions, we hope to show that translating from one solution to the other is not challenging. 

$\bullet$ We present a simple implementation for computing properties of network graphs in cuDF and show speedups of 147x-509x on an NVIDIA A100 GPU, 243x-1269X for an NVIDIA H100 GPU, and 332X-2185X for an NVIDIA H200 GPU over the single-core CPU implementation running in Pandas \cite{mckinney2011pandas}. In comparison to the reference GraphBLAS solution used in the verification phase, our algorithm is up to 30x faster than a multi-threaded version of the SuiteSparse GraphBLAS implementation running on a high-end CPU processor.

\section{Related Work}
\label{sec:related}
The MIT/Amazon/IEEE Graph Challenge is a suite of problems designed to support research in computational graph analysis. Similar to Top500 \cite{dongarra1997top500}, Graph500 \cite{graph500}, Kaggle competitions (\texttt{\url{https://www.kaggle.com}}), and the DARPA Grand Challenges (\texttt{\url{https://www.darpa.mil}}), the Graph Challenge provides problems with a precise definition without prescribing the solution approach. This encourages research activity by providing a common reference point across which to compare algorithmic, implementation, and hardware innovations. The Graph Challenge, at the time of writing, consists of four challenges involving static graphs \cite{samsi2017static}, streaming graphs \cite{kepner2020multi}, sparse deep Neural Networks \cite{kepner2020graphchallenge}, and the new anonymized network sensing challenge \cite{hayden2024}.

The work described in this paper focuses on the newly announced Anonymized Network Sensing problem \cite{hayden2024}. The Challenge involves loading a large set of PCAP data, creating a sparse matrix based on IP address connections, IP anonymization, and basic analysis of the resulting graph to compute network properties.

\paragraph*{\bf GraphBLAS}
The GraphBLAS \cite{davis2019algorithm} is an API specification that defines several fundamental building blocks of graph analysis in terms of sparse matrices and linear algebra operations.  The stable, domain-specific interface allows graph analyses to be compactly specified in a manner that retains mathematical context and remains orthogonal to implementation specifics. The queries in this challenge are naturally and compactly expressed in GraphBLAS, and it has been shown to be effective for performing the analyses in this challenge.  However, the ETL (extract, transform, load) and anonymization steps in the Graph Challenge are not directly expressible, and other approaches must be taken for these steps.  

\paragraph*{\bf Hypersparse Data} Sparse data sets can be found in many domains. This has resulted in the development of several unique sparse data representations such as CSR, CSC, COO, and ELL. The need for different structures has to do with the distribution of the data. The data set in this graph challenge highlights the significance of the hypersparse distribution, where many rows have few non-zero values, many rows are empty \cite{kepner2021mathematics}. 

\paragraph*{\bf Data Science - Dataframes and Python}
Over the last fifteen years, libraries such as Pandas \cite{mckinney2011pandas} and NumPy \cite{oliphant2006guide} have become increasingly popular within the data science and Python communities. The DataFrame data structure offers ease of use and robust functionality that includes sorting, joins, and groupbys across multiple columns in tabular data and has other functionality needed for ETL in machine learning and other data analysis applications. While customized solutions can sometimes offer better performance over commodity solutions, the extensive functionality of data science libraries makes them turnkey solutions. 

Integration into the larger Python ecosystem happens through the use of standard mathematical frameworks such as \texttt{NumPy}  which offer interoperability across hundreds of libraries. Many of these solutions are foundational; they are also inherently sequential, limiting their applicability to larger problems and systems. RAPIDS cuDF is a GPU variation of the Pandas library that targets NVIDIA GPUs. Similarly, cuPy \cite{nishino2017cupy} is a GPU extension of \texttt{NumPy} and cuGraph is an extension of NetworkX ~\cite{hagberg2008exploring}. Many of the CPU and GPU frameworks are interoperable and have similar APIs, but these may vary in some cases to better utilize the underlying hardware. In this paper, we target cuDF and show that we can implement the graph challenge without the need to explicitly use a graph library.

\section{Graph Challenge Setup}\label{sec:experiments}
\begin{table}[tb]
\small
\vspace*{-0.1 cm}
\caption{GPUs and CPUs used in our experiments.}

\scriptsize
\centering
\vspace*{-0.1 cm}

\begin{tabular}{cccccccc} 
\toprule
\textbf{GPU} & \textbf{SM} &  \textbf{CUDA} & \textbf{DRAM Size}  & \textbf{Power} & \textbf{Type}   \\ 
& &  \textbf{Cores} & \textbf{\& Type}  & &   \\ 
\midrule
A100 (Ampere) & 108    & 6912 & 80GB / HBM2e & 400W & SXM \\
H100 (Hopper) & 132  & 16896 & 80GB  / HBM3 & 350W & PCI-e \\ 
H200 (Hopper) & 132   & 16896 & 144GB /HBM3e & 700W & SXM \\ 
\bottomrule
\end{tabular}

\begin{tabular}{cccccc}
\toprule
\textbf{CPU} & \textbf{Processor} & \textbf{Cores} & \textbf{Threads} & \textbf{Sockets} & \textbf{DRAM Size} \\ 
& & & & & \textbf{\& Type}   \\
\midrule
x86-64   & Intel Xeon & 56 & 112 & 2 & 2TB \\  
& Platinum 8480+ & & &  & DDR5-4400  \\  
\bottomrule
\end{tabular}

\label{tab:gpu-cpu-systems}
\end{table}

\paragraph*{\bf Software and Libraries}
One of the goals of this work is to demonstrate that the graph challenge can be implemented using commodity hardware and software while achieving good performance. While some components of the challenge, especially the verification phase, use explicit GraphBLAS formulation, other parts are left open to interpretation. We implemented all phases of this graph challenge using the NVIDIA RAPIDS ecosystem \cite{RAPIDS}. The NVIDIA RAPIDS ecosystem includes numerous packages including cuDF, cuML, cuGraph, cuSpatial, and many more. Of these, we only required functionality from cuDF. 

An obvious question is: why did we not use cuGraph for our implementation? The answer is that we did consider using cuGraph initially; however, on deeper inspection of the graph challenge we saw that we could implement all the functionality using a traditional data-science approach and explored this path instead. Beyond that, we realized that several key components of the graph challenge focused on ETL components; as such, we decided to focus on a data science only solution. Using cuGraph is a topic of interest for future work. Our implementation uses the RAPIDS 25.02 release and is available for download as a Docker container via the NVIDIA NGC Catalog (\texttt{\url{https://catalog.ngc.nvidia.com/}}) or GitHub (\texttt{\url{https://github.com/rapidsai}}).

\paragraph*{\bf Hardware Configurations}
Table \ref{tab:gpu-cpu-systems} presents the GPUs and hardware details used in our experiments. We used GPUs from two different generations: the NVIDIA A100 GPU from the Ampere line of GPUs and the NVIDIA H100 and H200 from the Hopper line of GPUs. These GPUs included both the PCI-E and the SXM form factors, the latter of which supports fast NVLINK connection when run in a multi-GPU environment. Our current solution is single-GPU-based, and once the data is on the GPU, there is no data movement between the CPU and GPU. As such, the form factor is not critical, but we capture it for the sake of completeness.

Each of the aforementioned GPUs were part of a different server and connected to a different CPU, network adapters, and a different generation of PCI-E. In addition, the GPUs and their respective servers could be connected to the network file system via different networking switches. Together, these elements have obvious impacts on the performance of the benchmark. After the initial data-transfer from the file-system to the device, these hardware components no longer impact the performance of the application. As such, the only CPU specs that we provide in Table \ref{tab:gpu-cpu-systems} is based off the system we used for benchmarking the CPU-only implementations (the Graph Challenge reference code and our new algorithm running with Pandas as the backend). The first time that the graph is read to the server with the GPU, the file needs to be copied over the networked file system (across the switch and NIC) -- in the Graph Challenge specifications, this is referred to as the non-cached run. On the second run, the file is already cached in the memory of the CPU server (this is an artifact of the Linux operating system). This is referred to as a cache run.  Note, in the performance section of the new Graph Challenge (Jananthan \emph{et al.}), there is a differentiation between the cached and non-cached runs similar to the one highlighted above. There are substantial runtime differences between the non-cached and cache version for our new implementation. These are less impactful for the reading of data in the PCAP and GraphBLAS formats, see Table \ref{tab:io_time}. Lastly, the non-cached and cached versions mostly impact the data load time and do not impact the time of the computational phase which was the main focus of our work.

\paragraph*{\bf Inputs} At the time of writing, there is only a single input graph available for the new graph challenge. This graph is made up of $2^{30}$ edges and is available for download in two formats: PCAP and GraphBLAS ready. The PCAP file in its uncompressed format is roughly 67GB of data in comparison to the GraphBLAS formatted version, which is 8.1 GB of data.

\section{Implementation}
\label{sec:implementation}
Our proposed solution to the Network Sensing Graph Challenge follows the Challenge specification in~\cite{hayden2024}: A data loading stage that reads the dataset(s) in either PCAP or GraphBLAS format, parses and constructs the graph; a second step anonymizes the IP addresses contained in the graph, and writes the data to a binary file in \texttt{Parquet} format; and the final stage uses a data science approach to compute the graph properties specified in~\cite{hayden2024}. All steps are implemented using commodity and open-source software. Our solution did not rely on customized code for performance. 

Conceptually, the data reading and parsing steps can be implemented together in a true data science pipeline. We separate them in this paper for two reasons: 1) they use different parts of the dataframe ecosystem, and 2) we time them separately to capture the time costs of the storage, network, and data loading vs. the cost of data cleaning.

\paragraph*{\bf Data loading and parsing}
\label{subsubsec:loading}
The input data for this challenge is provided in two formats: PCAP and a \texttt{tar} archive containing multiple \texttt{tar} files, each with multiple GraphBLAS matrices. We developed a data loading pipeline for both formats and also saved the data in the \texttt{Parquet} format for fast reads when focusing on the graph analytics part of this challenge. Both these data loading pipelines use entirely open-source and commodity libraries with their own performance tradeoffs -- with PCAP parsing having scaling limitations and the \emph{tar} based solution being CPU bound. While being CPU-bound might not be a problem for some solutions, our goal was to show a solution that can be entirely accelerated with the GPU -- and that goal can be attained if the data is stored in a more GPU-friendly format.  Therefore, we also stored the input in \texttt{Parquet} and \texttt{csv} (comma separated values) format and found that these can be loaded more efficiently, in addition to being more user-friendly.

\paragraph*{\bf PCAP Format And Limitations}
Network data is predominantly available in the \texttt{PCAP} format. Thus, we used an open-source, Python library, \texttt{dpkt}~\cite{dpkt}, for reading and parsing the provided \texttt{PCAP} file. Additionally, we used the \texttt{PCAP-parallel}~\cite{pcap-parallel} package to enable a parallel, multi-threaded reading pipeline for \texttt{PCAP} data. This is an open-source Python package that uses standard multi-processing capabilities in Python to split large \texttt{PCAP} file reads across multiple cores/threads. In addition to this, we developed a multi-threaded approach to reading the same data from the GraphBLAS \texttt{tar} file distributed as part of this challenge. 

cuDF supports file IO (both reads and writes) of CSV (Comma Separated Values), Parquet, and a few additional formats. While some CPU data science solutions support PCAP parsing, cuDF does not natively support the parsing of PCAP files. 
Having said that, we emphasize that having support does not ensure performance, as we found with the native CPU solutions. Also, with the desire of making PCAP data more accessible, people should consider storing the data in a more data science friendly format.

We started with the CPU Python PCAP \texttt{PCAP-parallel} \cite{pcap-parallel} and saw that the average time to parse the file took roughly 2500 seconds with 256 CPU threads. 

While we could have used more threads, we found the single-core performance below expectations, and thus, even with perfect linear-performing, the PCAP reader with larger thread counts would perform poorly in comparison to faster parsing available in RAPIDS cuDF.

To ensure good performance and minimize CPU usage in our implementation, we stored only the necessary fields from the PCAP data into a \texttt{Parquet} file. 
With the data readily available in Parquet format, we could easily use cuDF to read the data directly into the GPU's memory. While the read times varied, we found average read times for reading from Parquet files using cuDF of the order of 14-22 seconds across all the GPUs used in this paper.

For all the tested systems, we ran the benchmark a second time for the Parquet input such that the Parquet file was cached by the operating system. Jananthan \emph{et al.} \cite{hayden2024}  refer to this as the cached-based solution, and this removes any bias of external features such as the network, the network adapter, or the underlying file storage solution. 

Given the scope of the paper, we limited our benchmarking of these features, but we do not dismiss them by any means. Further, we emphasize the significance in choosing the optimal data format for storing the PCAP data as a graph. While we understand that the PCAP format has additional flexibility over other formats, specifically for network data, we believe that the entire data set can be stored in either Parquet or a CSV format when it is captured -- making it more data science friendly. 

\paragraph*{\bf GraphBLAS format} In this format, the data is organized in a multi-tier hierarchy of \texttt{tar} files, with 128 \texttt{tar} files in the next level, each of which contains 64 matrices in GraphBLAS format. The fine-grain partitioning of the data allows for fine-grain access for large multi-threaded CPUs. In contrast, this representation is a performance hindrance for GPUs, as the data is loaded in small chunks of which would lead to GPU under-utilization. Thus, we did not explore a GPU variation for this reader.

For this implementation, we also used multiple threads for data reads. The read times for each approach are shown in Table~\ref{tab:io_time}. We observed an order of magnitude slower read time for the \texttt{PCAP} format approach compared with the  \texttt{tar}/GraphBLAS approach. We attribute this to the additional overhead of parsing individual packets from the \texttt{PCAP} format. 

Once the data is read into memory and the graph is constructed, we also save the network graph as a \texttt{Parquet} file, which leads to significantly improved read times (shown in Table~\ref{tab:io_time}).

In our initial analysis of the PCAP and GraphBLAS data ingest, we used a dual AMD EYPC 7742 processor with 128 cores and 256 threads. 
Due to the relatively low performance of the PCAP reader in both sequential and multithreaded modes, we chose not to rerun benchmarks with the PCAP file reader, and used the GraphBLAS matrices and \texttt{Parquet} formatted data for the rest of the challenge.

\begin{table}[t]
    \centering
    \caption{Non-cached and cached data read times were measured with 256 threads for \texttt{PCAP} and GraphBLAS formats. \texttt{Parquet} reads used a single \texttt{cuDF} thread, including GPU transfer time. H200 benchmark timings are shown for the GPU execution.}
    \label{tab:io_time}
    \begin{tabular}{lcccc}
    \toprule
     & \textbf{PCAP} & \textbf{GraphBLAS} & \textbf{Parquet} & \textbf{Parquet} \\
     & \textbf{CPU} & \textbf{CPU}  & \textbf{Pandas, CPU} & \textbf{cuDF, GPU} \\
     \midrule
     Non-Cached & 2562.74  & 22.14  & 50.67  & 14.75 \\
     Cached     &    -     &   -    & 6.07   & 0.49 \\
     \bottomrule    
     \end{tabular}
\end{table}

\begin{table*}[h]
\centering
\caption{Comparison to Table 1 from Jananthan et al.  ~\cite{hayden2024}, where $A_t$ is the traffic matrix. The first four operations affect the whole matrix; the last five connect source to destination vertices. For reverse operations simply replace `src` and `dst`.
}
\label{tab:graph_operations}    
\begin{tabular}{lccl}
\toprule
Description & Matrix notation & Summation notation & Data Science implementation\\
\midrule
Valid packets & $\Sigma_{i}\Sigma_{j} A_t(i,j)$   & $1^{T} A_t 1$  & \texttt{df[\textquotesingle n\_packets\textquotesingle].sum()} \\
Unique links & $\Sigma_{i}\Sigma_{j} |A_t(i,j)|_{0}$ & $1^{T} |A_t|_0 1$  & \texttt{df[[\textquotesingle src\textquotesingle, \textquotesingle dst\textquotesingle]].drop\_duplicates().size} \\
Link packets from $i$ to $j$ & $A_t(i,j)$ & $A_t$ & \texttt{df.groupby(by=[\textquotesingle src\textquotesingle, \textquotesingle dst\textquotesingle]).value\_counts()}\\
Max link packets $d_{max}$ & $max_{ij}A_t(i,j)$ & $max(A_t)$  & \texttt{df.groupby(by=[\textquotesingle src\textquotesingle,\textquotesingle dst\textquotesingle]).value\_counts().max()} \\
Unique sources &  $\Sigma_{i} |\Sigma_{j} A_t(i,j) |_{0}$   & $1^{T} |A_t 1|_{0}$ & \texttt{df[\textquotesingle src\textquotesingle].unique()} \\
Packets from source $i$ &  $\Sigma_{j} A_t(i,j)$   & $A_t 1$ &  \texttt{df.groupby(by=[\textquotesingle src\textquotesingle])} \\
Max source packets $d_{max}$ & $ max_{i}\Sigma_{j} A_t(i,j)$ & $max(A_t 1)$&  \texttt{df.groupby(by=[\textquotesingle src\textquotesingle]).size().max()} \\
Source fan-out from $i$ & $\Sigma_{j} |A_t(i,j)|_{0}$ & $|A_t|_{0} 1$ &  \texttt{df[[\textquotesingle src\textquotesingle]].value\_counts()} \\ 
Max source fan-out $d_{max}$ & $max_{i}\Sigma_{j} |A_t(i,j)|_{0}$ & $max(|A_t|_{0} 1)$ &  \texttt{df[[\textquotesingle src\textquotesingle]].value\_counts().max()} \\

\bottomrule
\vspace{-.25cm}
\end{tabular}
\end{table*}

\paragraph*{\bf IP Address Anonymization}

We use a simple, scalable approach for anonymization of the IP addresses contained in this dataset. Our approach is as follows: Given an array of length ${N}$, we generate the sequence array with the values: ${0,1,.. N-1}$, followed by the use of the Python \texttt{shuffle} operation. Using the \texttt{cupy.random.shuffle} operation with the sequence array generates a random permutation. 
The shuffled array is then used to assign new values to the original source and destination IP addresses using gather instructions. 

The size ${N}$ is data dependent and determined by the number of unique value of $src$ and $dest$ ids in the PCAP file. Namely, we need a unique ID for each source and destination IP address in the input. The \texttt{unique} function is provided by dataframe libraries and is used for a wide range of applications, including for removing duplicate rows. While we are not interested in the exact implementation of \texttt{unique} we note that two ways to implement this would be with a sort function (and finding elements that are non repetitive) or by using a hash-table and finding non-overlapping operations. 

The likelihood that the \texttt{shuffle(N)} operation returns the same input sequence, twice in a row, is very small $1/|N|!$. Thus, it is safe to assume that the order of the original sequence is not the same as the output allowing for good anonymization. To further ensure the uniqueness of the permutation, one can run the shuffle operation multiple times in a row on the generated sequence. From a performance standpoint, the shuffle operation is computationally demanding; however, one or two extra iterations do not greatly impact the overall performance given the overheads of the \texttt{gather} and \texttt{unique} operations.
Furthermore, the random permutation generator algorithm by Green \emph{et al}. \cite{green2022generating} could further reduce this cost as it was proven to be faster than the CuPy \texttt{shuffle} operation. The HashGraph \cite{green2021hashgraph} based permutation \cite{green2022generating} also enables deterministic testing (if necessary). However, we did not use this algorithm as our goal was to use commodity software.

\paragraph*{\bf Graph Analytics}
\label{subsec:analytics}
The final step in this challenge consists of computing graph properties described in Jananthan \emph{et al.} ~\cite{hayden2024} and Kepner \emph{et al.}  ~\cite{kepner2020multi}. Here, we leverage the RAPIDS cuDF package to compute these properties on the GPU. In our implementation, we represent the network graph as a cuDF DataFrame object with three columns: one corresponding to the source, one for the destination IP address, and lastly one column for the packet count between each source and destination IP address. The rest of the section will refer to these columns as \texttt{src}, \texttt{dst}, and \texttt{n\_packets} respectively. This allows us to leverage GPU-optimized implementations of various DataFrame operations to quickly compute the graph properties. Table \ref{tab:graph_operations} compares our new data science-based solution with the GraphBLAS-based formulation of the same operations. The graph operations are defined in Table 1 of the Graph Challenge benchmark ~\cite{hayden2024}. 

Globally unique IP addresses are trivial to find by using the \texttt{unique()} method of the DataFrame on each column and subsequently removing common values across the two unique sets. For computing other properties of the graph, we can compose multiple operations sequentially. 

Other properties of the graph are easily calculated using combinations of the \texttt{groupby()}, \texttt{value\_counts()}, \texttt{unique()}, and other commands shown in Table~\ref{tab:graph_operations}.

\textit{Unique links} – Finding all unique links in the dataset is achieved simply by using the \texttt{drop\_duplicates} method, followed by the \texttt{size} method:

\begin{lstlisting}[language=Python]
df[['src', 'dst']].drop_duplicates().size()
\end{lstlisting}

    \textit{Link packets from \textit{i} to \textit{j}} - We use the \texttt{groupby()} method to first determine all connections between source IP \textit{i} and destination IP \textit{j}. Once all links between \textit{i} and \textit{j} are found, the number of packets between each pair is simply a matter of calling the \texttt{value\_counts()} method. The \texttt{value\_counts()} function gives a count of the number of occurrences of each source and destination pair of IP addresses in the dataset. After this collective operation, we can trivially compute aggregations such as \texttt{min, max}, etc. to extract the desired graph properties. This can be simply written as the following command :

\begin{lstlisting}[language=Python]
df.groupby(by=['src', 'dst']).value_counts()
\end{lstlisting}

    \textit{Packets from source \textit{i}} - These are calculated using the \texttt{groupby()} operation on the \texttt{src} column as follows : 
\begin{lstlisting}[language=Python]
df.groupby(by=['src'])
\end{lstlisting}

\textit{Max source packets $d_{max}$} - This follows from the above \texttt{groupby()} with the use of the \texttt{size()} method and the aggregation function \texttt{max()} : 
\begin{lstlisting}[language=Python]
df.groupby(by=['src']).size().max()
\end{lstlisting}

\textit{Max source Fan-out} - Fan out is the number of connections from a given source. Thus, we can once again use the \texttt{value\_counts()} function to get a count of the number of occurrences of each source IP address, followed by a \texttt{max()} operation on the result, to get the maximum fan-out i.e. the maximum number of network connections observed from all sources as follows: 

\begin{lstlisting}[language=Python]
df[['src']].value_counts().max()
\end{lstlisting}

The same sequence of operations performed on the network destination column gives us the maximum fan-in value as well as the number of connections to a particular destination IP address. All of these operations are listed in Table~\ref{tab:graph_operations} along with the equivalent formulations in linear algebra and summation notation. The actual compute performance for computing properties based on the network source and destination will depend on the cardinality of each column and is the subject of future work on real world data. Finally, we also validated the results of the network analysis by comparing the results with the expected values obtained from the Challenge authors~\cite{hayden2024}.

\section{Productivity \& Performance Analysis}
\label{sec:results}

\paragraph*{\bf Data Science approach}
In Section ~\ref{sec:implementation}, we show a data science approach to network data analysis. A closer look at the implementation in Table ~\ref{tab:graph_operations} reveals that a few core operations in cuDF (and Pandas) are sufficient to perform all the graph analytic operations prescribed in this Graph Challenge. These operations can be summarized into filtering operations such as grouping, sorting, finding unique values, and aggregation/collectives such as sum, max, min, etc. By implementing the network analysis operations in basic cuDF/Pandas operations such as \texttt{groupby}, \texttt{value\_count} and \texttt{drop\_duplicates}, network data analysis becomes more accessible to a wider field of data scientists who can quickly and easily use tools and approaches they are already familiar with, to perform advanced analytics on network graphs. This simple re-formulation of more complex linear algebra routines also makes the solution more accessible to non-experts and enables rapid experimentation on large-scale network data.

\subsection*{Empirical Analysis}

\paragraph{\bf cuDF comparison with Pandas}
We compare the performance of our solution running on the GPU with cuDF and on the CPU with Pandas. The implementations are algorithmically identical, at least code-wise. However, the underlying implementations by the data science frameworks can be different -- which can also impact the performance.

We start off with the performance of the verification kernels of the Graph Challenge, as the results of these functions are independent of the implementation and should be the same for any Graph Challenge Solution. We then compare some of the pre-query phases that were part of the anonymization phase.

\begin{figure}[t]
    \centering
    \includegraphics[width=.97\linewidth]{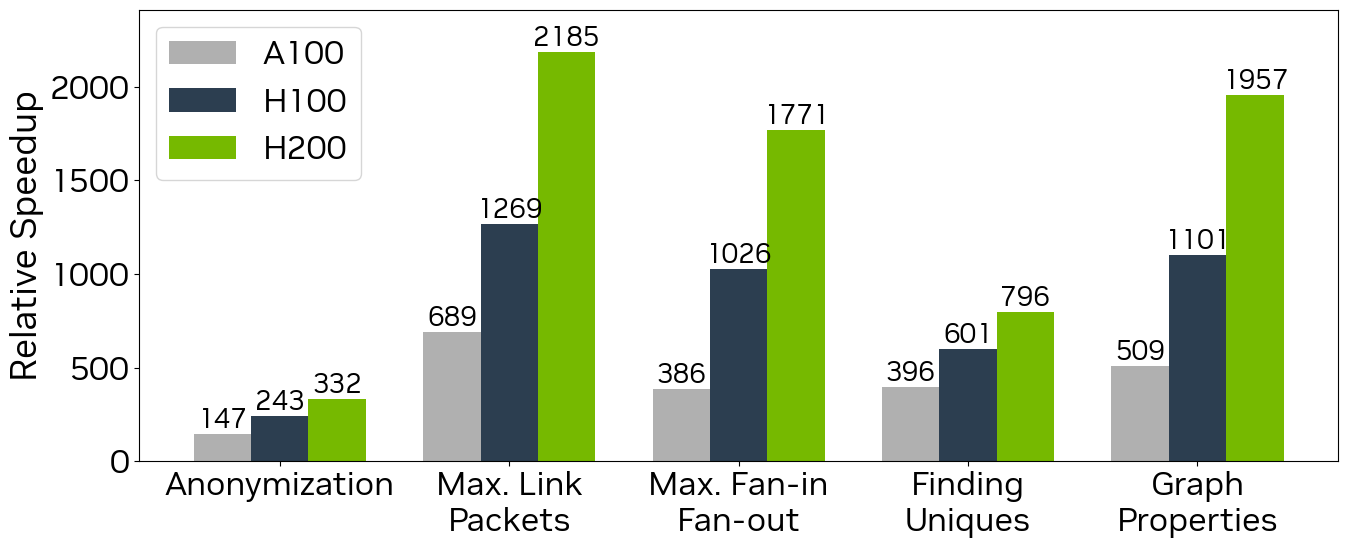}
    \caption{Comparing performance of cuDF implementation with the single-core Pandas implementation. Here, the \textit{Graph Properties} speedup is calculated by including the time needed to compute (max) fan-in/fan-out, finding unique source and destination IPs, max link packets, and the number of unique links.}
    \label{fig:all-pandas-speedup}
\end{figure}

Fig. \ref{fig:all-pandas-speedup}
depicts the speedup of our implementation running on the GPU with cuDF over its CPU counterpart running on sequential Pandas executing the same code. The speedups vary quite a bit depending on the function, as well as the GPU. We start off with the key observation, that for all the tested components, the performance increases from the NVIDIA A100 to the NVIDIA H100 and up to the NVIDIA H200 GPU. For the sake of brevity, we will focus on the performance analysis of the H200 GPU.

We start off with several of the faster queries, finding the maximal fanout and maximal number of edges are accelerated to 1771 and 2185x, respectively. The process of finding unique values is accelerated by a factor of 796x. In addition to this operation being necessary in the property query phase, we also used this function in the anonymization process.
These operations benefit from the GPU's higher bandwidth and possibly algorithmic optimization within their respective implementations. 
Together, our implementation using cuDF is 1957x faster than its Pandas counterpart to calculate all 14 queries in Table \ref{tab:graph_operations}. 

Lastly, we take a look at how much faster our anonymization pipeline is on the GPU in comparison to running on the CPU with Pandas. This analysis does not include the time it takes to load and clean the data from storage, as these systems have different storage systems and network cards. We refer the reader back to the analysis earlier in this paper on the difference between file formats with non-cached and cached execution times. Our anonymization pipeline is 392x faster on the GPU - this includes the time for finding unique IP addresses, generating a permutation, shuffling the data, and more. We found that the permutation generating operation was one of the more costly phases, and believe that it can be further accelerated. Nonetheless, the overall speedup was more than satisfactory as our starting point.

\begin{figure}[t!]
\centering
    \includegraphics[width=.97\linewidth]{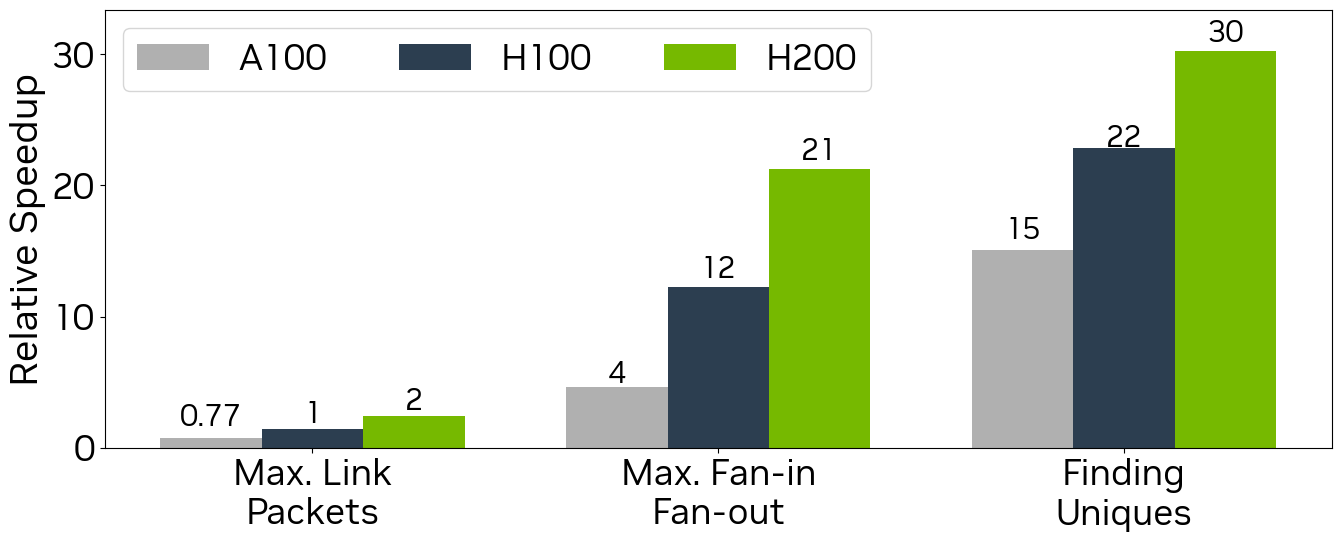}
    \caption{Comparing performance of cuDF running on a single GPU with multi-threaded GraphBLAS, using the reference implementation, running on an entire CPU socket.}    
    \label{fig:graphblas-speedup}
\end{figure}

\paragraph{\bf cuDF comparison with GraphBLAS}
The second implementation we compare against is the \textit{\textbf{reference implementation}} provided by the Graph Challenge. That implementation is OpenMP-based and runs on all the cores and threads available in the system (shown in Table \ref{tab:gpu-cpu-systems}). 
Fig.~\ref{fig:graphblas-speedup} shows the relative speedup of our implementation running on a {\bf single} GPU with cuDF over a multi-core implementation using SuiteSparse GraphBLAS ~\cite{davis2019algorithm} version 8.2.1 running on all cores on the node. 
When counting unique IP addresses, the A100 is over 15X faster than its CPU counterpart, whereas the H200 is close to 30X faster. For the fan-in and fan-out computations, the speedups are again lower (for similar reasons), with the A100 and H200 being 4X and 21X faster, respectively.
For the last of the property queries, GraphBLAS is fairly competitive, and we need to explore why that is the case. It is quite likely that GraphBLAS is storing the number of edge counts as the edge weight, allowing for a simple scan of the edges versus our more complex group-by operation over a multi-graph.

\section{Conclusions}
\label{sec:conclusions}

In this paper, we presented a solution for the anonymized network sensing graph challenge. Our goal was to use commodity hardware with NVIDIA A100, H100, and H200 GPUs. From a software perspective, our goal was to implement a solution using only off-the-shelf software without implementing any custom compute kernels. We achieved the latter of these goals using the RAPIDS ecosystem primarily using cuDF and cuPY. While the Graph Challenge does not formulate the graph building process, the verification process uses the language of the GraphBLAS. Our solution is entirely based on a data science centric approach using open-source tools. We also showed how to formulate the GraphBLAS queries with one-line data science code. Thus, we also closed the gap between data scientists and GraphBLAS based applications.
Lastly, from a performance perspective, we showed that our solution running on commodity hardware is significantly faster than the \textit{\textbf{reference implementation}} based on GraphBLAS. We showed that our solution can execute using the CPU Pandas data science stack, though from a performance perspective, our solution greatly benefits from running on the GPU as the various kernels can fully exploit the available parallelism, allowing for kernel speedups in the range of 204X-2185 using a single NVIDIA H200 GPU.

\balance
\bibliographystyle{IEEEtran} 
\bibliography{IEEEabrv, bibfile}

\end{document}